\documentclass[11pt,a4paper]{article}
\usepackage{times,latexsym}
\usepackage{xurl}
\usepackage[T1]{fontenc}
\usepackage{multirow}
\usepackage{textcomp}
\usepackage{inconsolata}

\newcommand{\shorteq}{%
  \settowidth{\@tempdima}{-}
  \resizebox{\@tempdima}{\height}{=}%
}

    \usepackage[acceptedWithA]{tacl2021v1}
\usepackage{tacl2021v1}
\usepackage{xspace,mfirstuc,tabulary}

\newif\iftaclinstructions
\taclinstructionsfalse 
\iftaclinstructions
\renewcommand{\confidential}{}
\renewcommand{\anonsubtext}{(No author info supplied here, for consistency with
TACL-submission anonymization requirements)}
\newcommand{\instr}
\fi

\iftaclpubformat 

\else

\fi

\usepackage{xcolor}

\definecolor{myred}{RGB}{243, 171, 168}
\definecolor{myblue}{RGB}{171, 201, 234}
\definecolor{mygreen}{RGB}{153, 218, 167}
\definecolor{myorange}{RGB}{239, 183, 147}

\newcommand{\calC}{\mathcal{C}}

\newcommand{\calO}{\mathcal{O}}

\newcommand{\bo}{\mathbf{o}}

\newcommand{\longformer}{\mathrm{Longformer}}
\newcommand{\transformer}{\mathrm{Transformer}}
\newcommand{\BERT}{\mathrm{BERT}}
\newcommand{\LBERT}{\operatorname{LEGAL-BERT}}
\everypar{\looseness=-1}

\usepackage{amsfonts}
\usepackage{amsmath}
\usepackage{bbm}
\usepackage{amsthm}
\usepackage{times}
\usepackage{here}
\usepackage{latexsym}
\usepackage{booktabs}
\usepackage{svg}

\usepackage{graphicx, natbib}

\usepackage{subfig}
\usepackage{cleveref}
\usepackage{cancel}

\crefname{section}{\S}{\S\S}
\Crefname{section}{\S}{\S\S}
\crefname{table}{Table}{}
\crefname{figure}{Figure}{}
\crefname{algorithm}{Algorithm}{}
\crefname{equation}{Eq.}{}
\crefname{appendix}{Appendix}{}
\crefname{thm}{Theorem}{}
\crefname{prop}{Proposition}{}
\crefname{cor}{Corollary}{}
\crefname{observation}{Observation}{}
\crefname{assumption}{Assumption}{}
\crefformat{section}{\S#2#1#3}

\newtheorem{assumption}{Assumption}

\usepackage{comment}
\usepackage{todonotes}

\newcommand{\precedent}{Outcome Corpus}
\newcommand{\chalkidis}{\citeauthor{chalkidis-etal-2021-paragraph} Corpus}
\newcommand{\model}{claim--outcome model}
\newcommand{\corpus}{outcome corpus}

\newcommand{\R}{\mathbb{R}}

\newcommand{\softmax}{{\footnotesize \textsf{softmax}}}
\newcommand{\relu}{\rho}
\newcommand{\enc}{{\footnotesize \textsf{enc}}}
\DeclareMathOperator*{\argmax}{\footnotesize \textsf{argmax}}

\everypar{\looseness=-1}
\usepackage{tikz}
\newcommand{\Plus}{\mathord{\begin{tikzpicture}[baseline=0ex, line width=1, scale=0.13]
\draw (1,0) -- (1,2);
\draw (0,1) -- (2,1);
\end{tikzpicture}}}

\newcommand{\Minus}{\mathord{\begin{tikzpicture}[baseline=0ex, line width=1, scale=0.13]
\draw (0,1) -- (2,1);
\end{tikzpicture}}}

\usepackage{emoji}
\newcommand{\ucambridge}{\emoji[twitter_emoji]{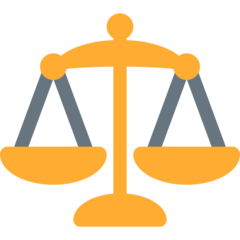}}
\newcommand{\eth}{\emoji[twitter_emoji]{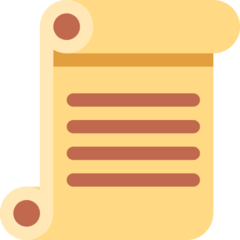}}

\title{On the Role of Negative Precedent in Legal Outcome Prediction}

\author{Josef Valvoda$^{\ucambridge}$ \quad Ryan Cotterell$^{\eth}$ \quad Simone Teufel$^{\ucambridge}$ \\
  $^{\ucambridge}$University of Cambridge \quad $^{\eth}$ETH Z\"urich \\
  \texttt{\{\href{mailto:jv406@cam.ac.uk}{jv406},\href{mailto:sht25@cam.ac.uk}{sht25}\}@cam.ac.uk} \\
  \texttt{\href{mailto:ryan.cotterell@inf.ethz.ch}{ryan.cotterell@inf.ethz.ch}}\\
}

\date{}

\begin{document}
\maketitle
\begin{abstract}

Every legal case sets a precedent by developing the law in one of the following two ways. 
It either expands its scope, in which case it sets positive precedent, or it narrows it, in which case it sets negative precedent. 
Legal outcome prediction, the prediction of positive outcome, is an increasingly popular task in AI. 
In contrast, we turn our focus to negative outcomes here, and introduce a new task of negative outcome prediction.
We discover an asymmetry in existing models' ability to predict positive and negative outcomes. 
Where the state-of-the-art outcome prediction model we used predicts positive outcomes at $75.06$ $F_1$, it predicts negative outcomes at only $10.09$ $F_1$, worse than a random baseline.
To address this performance gap, we develop two new models inspired by the dynamics of a court process. 
Our first model significantly improves positive outcome prediction score to $77.15$ $F_1$ and our second model more than doubles the negative outcome prediction performance to $24.01$ $F_1$. 
Despite this improvement, shifting focus to negative outcomes reveals that there is still much room for improvement for outcome prediction models.
\looseness -1

\vspace{0.5em}
\hspace{0em}\includegraphics[width=1.25em,height=1.25em]{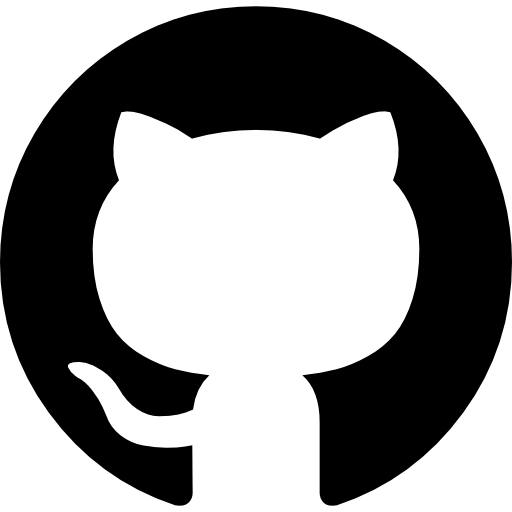}\hspace{.75em}\parbox{\dimexpr\linewidth-7\fboxsep-2\fboxrule}{\url{https://github.com/valvoda/Negative-Precedent-in-Legal-Outcome-Prediction}}
\vspace{-1.5em}
\end{abstract}

\section{Introduction}

The legal system is inherently adversarial.
Every case pitches two parties against each other: the \textbf{claimant}, who alleges their rights have been breached, and the \textbf{defendant}, who denies breaching those rights.
For each claim of the claimant, their lawyer will produce an argument, for which the defendant's lawyer will produce a counterargument.
In precedential legal systems \cite{garner2009black}, the decisions in the past judgements are binding on the judges in deciding new cases.\footnote{This is in contrast to the civil law jurisdictions, where judges do not create new law and predominantly rely on applying rules found in the Legal Code. Our paper mainly concerns precedential legal systems, such as the US, UK, Australia and India.}
Therefore, both sides of the dispute will rely on the outcomes of previous cases to support their position \citep{duxbury_2008, sep-legal-reas-prec, garner2009black}.
The claimant will assert that her circumstances are alike those of previous claimants whose rights have been breached.
The defendant, on the other hand, will allege that the circumstances are in fact more alike those of unsuccessful claimants.
The judge decides who is right, and by doing so establishes a new precedent.
If it is the claimant who is successful in a particular claim, the precedent expands the law by including the new facts in its scope.
If it is the defendant who is successful, the law is contracted by rejection of the new facts from its scope.
The expansion or contraction is encoded in the case outcome; we will refer to them as \textbf{positive outcome} and \textbf{negative outcome}, respectively.

\looseness=-1


Positive and negative outcomes are equally binding, which means that the same reasons that motivate the research of the positive outcome also apply to the negative outcome. 
Both are important for computational legal analysis, a fact that has been known at least since \citet{lawlor}.\footnote{He describes them as \emph{pro-precedent} and \emph{con-precedent}.}
However, the \textit{de facto} interpretation of precedent in today's legal NLP landscape focuses only on positive outcomes. 
Several researchers have shown that a simple model can achieve very high performance for such formulation of the outcome prediction task \cite{aletras, chalkidis-etal-2019-neural, svm, lexGLUE}, a finding that has been replicated for a number of jurisdictions \cite{zhong-etal-2018-legal, xu2020distinguish}.\looseness=-1



In this work, we reformulate outcome prediction as the task of predicting both the positive and negative outcome given the facts of the case.
Our results indicate that while a simple $\BERT$-based classification model can predict positive outcomes at an $F_1$ of $75.06$, it predicts negative outcomes at an $F_1$ of $10.09$, falling short of a random baseline which achieves $11.12$ $F_1$.
This naturally raises the question: What causes such asymmetry?
In \cref{sec:discussion}, we argue that this disparity is caused by the fact that most legal NLP tasks are formulated without a deep understanding of how the law works.\looseness=-1

Searching for a way to better predict negative outcomes, we hypothesise that building a probabilistic model that is more faithful to the legal process will improve both negative and positive outcome prediction.
To test this hypothesis we develop two such models. 
Our first model, which we call the \textbf{joint model}, is trained to jointly predict positive and negative outcome.
Our second model, which we call the \textbf{\model}, enforces the relationship between the claims and outcomes.
While the joint model significantly\footnote{Throughout the paper we report significance using the two tailed paired permutation test with $p < 0.05$.} outperforms state-of-the-art models on positive outcome prediction with $77.15$ $F_1$, the \model{} doubles the $F_1$ on negative outcome prediction at $24.01$ $F_1$.
We take this result as
strong evidence that building neural models of the legal process should incorporate domain-specific knowledge of how the legal process works.
\looseness=-1

\section{The Judicial Process}\label{sec:precedent}

In order to motivate our two models of outcome, it is necessary to first understand the process of how law is formed.
Broadly speaking, the legal process can be understood as a task of narrowing down the legal space where the breach of law might have taken place. 
Initially, before the legal process begins, the space includes all the law there is, i.e., every legal Article.\footnote{Legal Article is a codification of a particular law. For example Article 3 of the European Convention of Human Rights prohibits torture.}
It is the job of the lawyer to narrow it down to only a small number of Articles, a subset of all law.
Finally, the judge determines which of the claimed Articles, if any, has been violated. 
We can therefore observe two distinct interactions between the real world and the law: (i) when a lawyer connects the real world and law via a claim, and (ii) when a judge connects them via an outcome.\looseness=-1

In practice, this means that judges are constrained in their decision.
They cannot decide that a law has been breached unless a lawyer has claimed it has.
A corollary of the above is that lawyers actively shape the outcome by forcing a judge to consider a particular subset of law.
In doing so a lawyer defines the set of laws from which the judge decides on the outcome.\footnote{In some jurisdictions and certain type of trials, the decision is made by a jury instead of a judge.}
The power of a lawyer is also constrained. 
On one hand, lawyers want to claim as many legal Articles as possible, on the other there are only so many legal Articles that are relevant to their client's needs.
Thus, there are two principles that arise from the interaction of a lawyer and a judge.
First, positive outcome is a subset of claims.
Second, negative outcome consists of the unsuccessful claims, i.e. the claims the judge rejected.\looseness-1

There is a close relationship between claims and negative outcomes: If we knew the claims the lawyer had made, we could define negative outcome as exactly those Articles that have been claimed, but that the judge found not to be violated.
Much like how outcomes are a product of judges, claims are a product of lawyers.
And, unlike facts, they are not known before human legal experts interact with the case.
Therefore, to study the relationship of outcomes and facts, one can not rely on claims as an additional input to the model.
The only input which is available and known before a case is processed by the court, are the facts.\looseness=-1

\paragraph{Outcome prediction task.}
Legal facts are the transcript of the judges description of what has happened between the claimant and the defendant.
Under the current formulation of the outcome prediction task,
models are trained to predict whether case facts correspond to a violation of each Article, i.e., the models are trained to predict a vector in $\{0,1\}^K$ where $1$ indicates a positive outcome and $K$ is the number of legal Articles under consideration.\looseness-1

\paragraph{What is wrong with current work?}
In the above formulation, $0$ is ambiguous, it can indicate either that the Article not claimed or that the judge ruled that that specific Article was not breached.
Existing models, which don't take any information about claims into accounts, implicitly assume that {\emph all} Articles have been claimed, which is almost never the case in practice.
Under this assumption, the role of the legal claim and of negative outcomes is therefore effectively ignored.

\paragraph{Reformulating the task.} 
How should negative outcome then be modelled?
Given the domain specific knowledge about the interaction of a judge and a lawyer, our position is that models that predict outcomes should model the claims and outcomes \emph{together}.
To this end, we first need information about which laws have been claimed.
In \cref{sec:corpora}, we discuss the creation of a new corpus which contains the necessary annotation for this task.
In the next section we develop two models that jointly predict outcomes and claims using two basic assumptions about how the law operates.
We believe that our reformulation of the task has two advantages. 
First, considering positive and negative outcomes \emph{together} is a step towards better evaluation of legal outcome prediction models. Second, incorporating the roles of a judge and a lawyer within the models of outcome is a step towards better models of law.

\begin{figure*}[ht]
\subfloat[\textbf{Simple baseline model}\label{fig:mod1}]
  {\includegraphics[width=.23\linewidth]{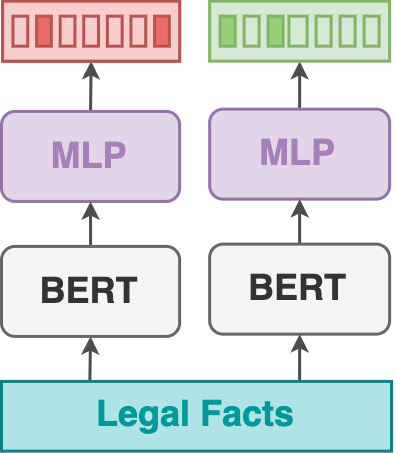}}\hfill
\subfloat[\textbf{MTL baseline model} \label{fig:mod3}]
  {\includegraphics[width=.23\linewidth]{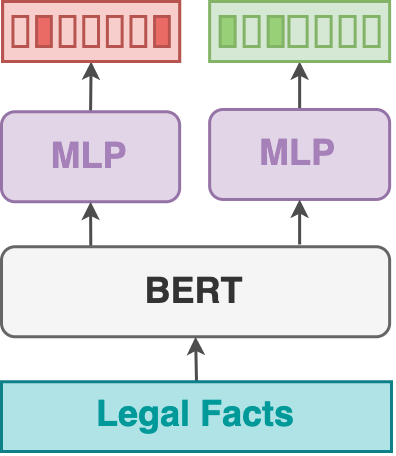}}\hfill
\subfloat[\textbf{Joint Model}\label{fig:mod4}]
  {\includegraphics[width=.23\linewidth]{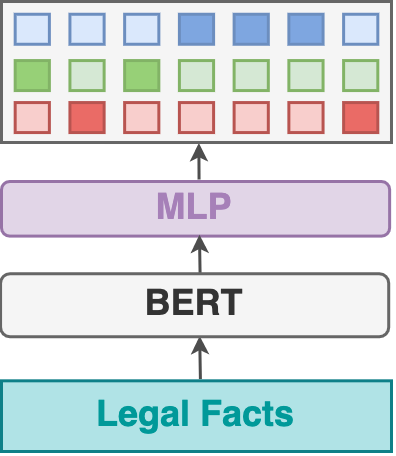}}\hfill
\subfloat[\textbf{Claim--Outcome Model}\label{fig:mod2}]
  {\includegraphics[width=.23\linewidth]{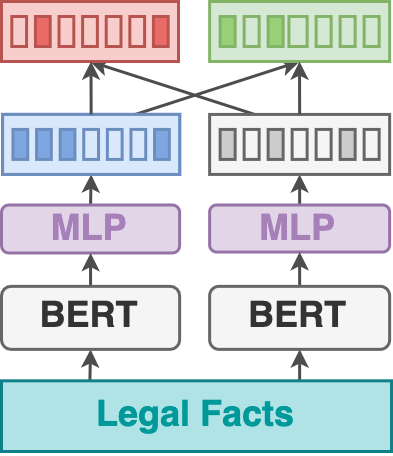}}

\caption{The models under consideration. Green and red boxes represent positive and negative outcomes respectively. Blue boxes represent claims. 
}
\end{figure*}

\newcommand{\Claim}{C}
\newcommand{\claim}{c}
\newcommand{\bclaim}{\mathbf{\claim}}

\newcommand{\positive}{+} 
\newcommand{\negative}{-}
\newcommand{\bpos}{\Plus}
\newcommand{\bneg}{\Minus}

\newcommand{\Fact}{F}
\newcommand{\fact}{f}
\newcommand{\bfact}{\mathbf{\fact}}
\newcommand{\W}{\mathbf{W}}
\newcommand{\U}{\mathbf{U}}
\newcommand{\V}{\mathbf{V}}
\newcommand{\F}{\boldsymbol{F}}

\newcommand{\equals}{=}

\newcommand{\mH}{\mathrm{H}}

\newcommand{\facts}{\boldsymbol{f}}
\newcommand{\outcomes}{\mathbf{o}}
\newcommand{\claims}{\mathbf{c}}
\newcommand{\outcomenull}{\emptyset}
\newcommand{\claimyes}{\textsc{y}}
\newcommand{\claimno}{\textsc{n}}
\newcommand{\bO}{\textbf{O}}
\newcommand{\bC}{\textbf{C}}
\newcommand{\mO}{\mathrm{O}}
\newcommand{\mC}{\mathrm{C}}
\newcommand{\mok}{\mathrm{O}_{k}}


\section{Law-Abiding Models}\label{sec:experiments}
In this section we formulate our two probabilistic models of law.
Our law-abiding models are built on top of the two assumptions described below.\looseness=-1
\paragraph{Notation.}
We define a probability distribution over three random variables.
\begin{itemize}
    \item $\mO$ is a random variable ranging over the set $\calO = \{+, -, \outcomenull\}$, whose elements correspond to positive, negative and null outcome, respectively. 
    The null outcome refers to all those Articles that the lawyer did not claim.
    The values of $\mO$ are denoted $o \in \calO$.
    We use a subscript, i.e., $\mO_k = o_k$, to refer to the random variable associated with the $k^{\text{th}}$ Article.
    Bolded  $\bO$ is a random variable ranging over $\calO^K$, where $K$ is the number of Articles we consider.
    The values of $\bO$ are denoted as $\bo \in \calO^K$.\looseness-1
    \item The random variable $\mC$ ranges over the set $\calC = \{\claimyes, \claimno\}$, whose elements encode whether or not an Article has claimed.
    The values of $\mC$ are denoted $c \in \calC$.
    We use a subscript, i.e., $\mC_k = c_k$, to refer to the random variable associated with the $k^{\text{th}}$ Article.
    Bolded $\bC$ is a random variable ranging over $\calC^K$.
    The values of $\bC$ are denoted as $\claims \in \calC^K$.\looseness=-1
    \item The random variable $\F$ ranges over textual descriptions of facts, i.e., $\Sigma^*$ for a vocabulary~$\Sigma$. 
    Values of $\F$ are denoted as $\facts$.\looseness-1
\end{itemize}

\subsection{Joint Model}
We begin with a simple assumption that, given the facts of a case, legal Articles are independent.
\begin{assumption}[Conditional Independence]\label{assumption1}
Conditioned on the facts $\F = \facts$, the random variables $(\mO_k, \mC_k)$ for the $k^{\text{th}}$ Article are jointly conditionally independent of the random variables $(\mO_{\ell}, \mC_{\ell})$ for the $\ell^{\text{th}}$ Article when $\ell \neq k$.
\looseness=-1
\end{assumption}
\noindent This assumption is based in the origin of each Article as an independent Human Right, related by the spirit of ECHR, but otherwise orthogonal in nature.
This is indeed how the law operates in general. 
A law, whether codified in an Article or a product of precedent, encodes a unique right or obligation.
In practice this means that a breach of one law does not determine a breach of another.
For example, a breach of Article 3 of ECHR (the prohibition of torture) does not entail a breach of Article 6 (right to a fair trial).
Even breaches of law that are closely related, for example libel and slander, do not entail each other, and allegation of each must be considered independently.

By \cref{assumption1}, the joint distribution over outcomes and claims decomposes over Articles as
\begin{align}
    p(\mathbf{O} &= \mathbf{o}, \mathbf{C} = \mathbf{c} \mid \F = \facts)  \\
    &= \prod_{k=1}^K p(\mathrm{O}_k = o_k, \mathrm{C}_k = c_k \mid \F =  \facts) \nonumber
\end{align} 
In the remainder of the text we write $\facts$ in lieu of $\F = \facts$ to save space.
We also write $o_k$ instead of $\mathrm{O}_k = o_k$ and $c_k$ instead of $\mathrm{C}_k = c_k$, respectively, when it is clear from context.\looseness=-1

\subsection{Claim--Outcome Model} 
Our second model builds on the first assumption with a second simple assumption:
\begin{assumption}[Claims and Outcomes]\label{assumption2}
For an Article to be breached, i.e., for it to become an outcome, it first needs to be claimed. The judge provides an outcome if and only if a claim is made:
\begin{subequations}
\begin{align}
    p(\mathrm{O}_k = \outcomenull \mid \mathrm{C}_k = \claimyes, \facts)  &= 0 \\
    p(\mathrm{O}_k = \outcomenull \mid \mathrm{C}_k =\claimno, \facts) &= 1
\end{align}
\end{subequations}
\end{assumption}
\noindent By \cref{assumption2}, we have that each distribution over outcome--claim pairs simplifies into the following equation:\looseness=-1
\begin{align}
    &p(\mathrm{O}_k = o_k, \mathrm{C}_k = c_k \mid \boldsymbol{f}) = \\
    \,\,\,&\begin{cases}
        p(o_k \mid \textsc{y}, \boldsymbol{f})\,   p(\textsc{y} \mid \facts) & \textbf{if } c_k = \textsc{y} \wedge	 o_k \neq \outcomenull \\
        p(\textsc{n} \mid \facts) & \textbf{if } c_k = \textsc{n} \wedge	 o_k = \outcomenull \\
        0 & \textbf{otherwise}
    \end{cases} \nonumber
\end{align}
Crucially, \cref{assumption2} allows us to reduce the problem to two independent binary classification problems.
First, we train a claim predictor $p(c_k \mid \facts)$ that predicts whether a lawyer would claim that the $k^{\text{th}}$ Article is relevant to the facts~$\facts$. 
Second, we train an outcome predictor that predicts whether the outcome is $+$ or $-$, \emph{given} that the lawyer has claimed a violation of Article $k$.
\looseness-1

\subsection{Neural Parameterization}\label{sec:models}
We consider neural parametrizations for all the distributions discussed above.
At the heart of all of our models is a high-dimensional representation $\enc(\facts) \in \R^{d_{1}}$ of the facts $\facts$. We obtain this representation from a pre-trained language model fine-tuned for our task (see \cref{para:plm}). 
All our language models rely on $\facts$ as their sole input. 
Except where we indicate otherwise, both the language model weights and classifier weights are learned separately for every model presented below.\footnote{For the language models, the weights are fine-tuned from a pre-trained model we initialise from the Hugging Face library.\looseness=-1}



\paragraph{Joint Model.}
First we parameterise the joint model which gives us a joint distribution over all configurations of $o_k$ and $c_k$ for a specific Article~$k$.
In principle, there are six such configurations $\{+, -, \outcomenull\} \times \{\textsc{y}, \textsc{n}\}$.
However, after we enforce \Cref{assumption2}, we are left with only three configurations $\Big\{\langle +, \textsc{y}\rangle$, $\langle -, \textsc{y}\rangle$, $\langle \emptyset, \textsc{n}\rangle \Big\}$.
This reduces the problem to a 3-way classification, which we parameterise as follows:\looseness=-1
\begin{align} 
    \label{eq:MTL_pos}
    p(\mathrm{O}_k = o_{k}, &\mathrm{C}_k = c_{k} \mid \facts) = \\
    &\softmax(\mathbf{U}_k\, \relu(\mathbf{V}_k\,\enc(\facts)))_{\langle o_k, c_k\rangle} \nonumber
\end{align}
where $\softmax(\mathbf{x})_i = \frac{\exp x_i}{\sum_{i'}^I \exp x_{i'}}$, $\relu$ is a $\mathrm{ReLU}$ activation function defined as $\relu(x) = \max(0, x)$; $\mathbf{U}_k \in \R^{3 \times d_2}$ and $\mathbf{V}_k \in \R^{d_2 \times d_1}$ are per-Article learnable parameters.
In total, the classifier has $K(3d_2 + d_2d_1)$ parameters, excluding those from the encoder $\enc$.

\paragraph{Claim--Outcome Model.}


We parametrise the \model{} as two binary classification tasks:
one which is predicting the claims, the other which is predicting positive outcomes. For the latter binary classification task, one class corresponds to $+$, while the other to both $-$ and $\outcomenull$.
This leads to the following pair of binary classifiers:
\begin{align} 
    \label{eq:MTL_pos}
    &p(\mathrm{C}_k = \claimyes \mid \facts) = \sigma(\mathbf{u}_{k} \cdot  \,\relu(\V_{k}\,\enc(\facts))) \\ 
    & p(\mathrm{O}_k = + \mid \mathrm{C}_k= \claimyes,\facts) = \sigma(\mathbf{u}'_{k} \cdot \,\relu(\V'_{k}\,\enc'(\facts))) \nonumber
\end{align}
where $\mathbf{u}_k \in \R^{d_2}$, $\mathbf{V}_k \in \R^{d_2 \times d_1}$, $\mathbf{u}'_k \in \mathbb{R}^{d_3}$, and $\mathbf{V}'_k \in \mathbb{R}^{d_3 \times d_1}$ are learnable parameters,
$\sigma(x)= \frac{1}{1 + \exp(-x)}$ is the sigmoid function, and $\enc$ and $\enc'$ are two \emph{separate} encoders.
In total, we have $K(d_2 + d_1d_2 + d_3 + d_1d_3)$ parameters, excluding those from the encoder $\enc$.
We use primed symbols to denote separately learned parameters. 
Given these probabilities, we can marginalise out the claims to obtain the probability of a positive outcome:\looseness=-1
\begin{align}  
    p(&\mathrm{O}_k = + \mid \facts)  \\
    &=p(\mathrm{O}_k = + \mid \mathrm{C}_k = \claimyes,\facts)\, p(\mathrm{C}_k = \claimyes \mid \facts) \nonumber\\ 
    &\quad + p(\mathrm{O}_k = +  \mid \mathrm{C}_k = \claimno, \facts)\, p(\mathrm{C}_k = \claimno \mid \facts) \nonumber \\
    &\stackrel{(1)}{=} p(\mathrm{O}_k = +  \mid \mathrm{C}_k = \claimyes,\facts)\, p(\mathrm{C}_k = \claimyes \mid \facts) \nonumber
\end{align}
where (1) is true because $p(\mathrm{O}_k = +  \mid \mathrm{C}_k = \claimno, \facts)$ is always zero (since by \cref{assumption2} no positive outcome can be set on an unclaimed case).

We then predict the probability of negative outcome as the complement of the probability of a positive outcome multiplied by the probability of a claim: 
\begin{align} 
    p(\mathrm{O}_k &= - \mid \facts) = \\
    &\Big(1-p(\mathrm{O}_k = +  \mid \claimyes,\facts) \Big)\, p(\mathrm{C}_k = \claimyes \mid \facts) \nonumber
\end{align}
This step enforces that the negative outcome probability is always both lower than that of claims and sums up to $1$ with the probability of positive outcome.
Finally, we have that
\begin{equation}
    p(\mathrm{O}_k = \emptyset \mid \facts) = p(\mathrm{C}_k = \claimno \mid \facts)
\end{equation}
To make a decision, we compute the following 
argmax that marginalises over claims:
\begin{align}
    o_k^\star &= \argmax_{o_k \in \calO} p(\mathrm{O}_k = o_k \mid \facts) \\
    &= \argmax_{o_k \in \calO} \sum_{c_k \in \mathcal{C}} p(\mathrm{O}_k = o_k, \mathrm{C}_k = c_k \mid \facts) \nonumber
\end{align}

\paragraph{Training and Fine-tuning.}
All models in \cref{sec:models} are trained by maximizing the log of the joint distribution $p(\mathbf{o}, \mathbf{c} \mid \mathbf{f})$.
We are given a dataset of triples $\mathcal{D} = \Big\{\left(\mathbf{o}^{(n)}, \mathbf{c}^{(n)}, \facts^{(n)}\right)\Big\}_{n=1}^N$
Due to independence assumption made, this additively factorises over Articles
$\sum_{n=1}^N \log p(\mathbf{o}^{(n)}, \mathbf{c}^{(n)} \mid \facts^{(n)})  = \sum_{n=1}^N \sum_{k=1}^K \log p(o^{(n)}_k, c^{(n)}_k \mid \facts^{(n)})$.
We fine-tune $\enc$ jointly for all $p(o_k, c_k \mid \facts)$.

\section{Baselines}

We contextualise the performance of the joint and \model{} with a number of baselines.
As a starting point we build a simple classification model trained to predict positive or negative outcome separately, 
see \cref{fig:mod1}.
We further want to test whether the advantage of our joint model stems from encoding the relationship between positive and negative outcome, or whether it is down to simply training on more data.
We test this by formulating the task as a multi-task learning objective, see \cref{fig:mod3}. While this model is trained on the same amount of data as our joint model, it does not explicitly encode the relationship between positive and negative outcomes.

\paragraph{A Simple Baseline.}
For our simple baseline model we formulate the positive and negative outcome prediction as a multi-label classification task.
Despite its conceptual simplicity, this model achieves state-of-the-art performance on the task of positive outcome prediction.
Given the facts of a case $\facts$, we directly model the probability that the outcome is positive, as a binary classification problem where the first class is the positive $+$ and the second is the union of negative and unclaimed $\{-, \emptyset\}$.
Likewise, we separately model the probability that the outcome is negative, as a binary classification problem where the first class is the negative $-$ and the second is the union of positive and unclaimed $\{+, \emptyset\}$.
To this end, we define a pair of binary classifiers:\looseness=-1
\begin{subequations}
\begin{align} 
    \label{eq:longformer}
    p(\mathrm{O}_k = + \mid \facts) &= \sigma(\mathbf{u}_{k}^{1} \cdot \relu(\V_{k}^{1}\,\enc_{1}(\facts))) \\ 
    p(\mathrm{O}_k = - \mid \facts) &= \sigma(\mathbf{u}_{k}^{2} \cdot \relu(\V_{k}^{2}\,\enc_{2}(\facts))) 
\end{align}
\end{subequations}
where $\mathbf{u}_k^{1} \in \R^{d_2}$, $\mathbf{V}_k^{1} \in \R^{d_2 \times d_1}$, $\mathbf{u}_k^{2} \in \mathbb{R}^{d_3}$, and $\mathbf{V}_k^{2} \in \mathbb{R}^{d_3 \times d_1}$ are the per-Article learnable parameters.
Thus, in total, we have $K(d_3 + d_3 d_1 + d_2 + d_2d_1)$ parameters excluding those from the fine-tuned encoders $\enc_{1}$ and $\enc_{2}$. 
The encoders $\enc_{1}$ and $\enc_{2}$ represent two \emph{different} fine-tuned parameters of the encoder.
Note that this approach does not model whether or not an Article is claimed, which stands in contrast to the main models proposed by this work.
\looseness-1



\paragraph{MTL Baseline.}
We also consider a version of the simple baseline where we jointly 
fine-tune a single encoder.
Symbolically, this is written as:
\begin{subequations}
\begin{align} 
    \label{eq:MTL_pos}
    p(\mathrm{O}_k = + \mid \facts) &= \sigma(\mathbf{u}_{k}^{1} \cdot \relu(\V_{k}^{1}\,\enc(\facts))) \\ 
    p(\mathrm{O}_k = - \mid \facts) &= \sigma(\mathbf{u}_{k}^{2} \cdot \relu(\V_{k}^{2}\,\enc(\facts)))
\end{align}
\end{subequations}
where $\enc$ is shared between the classifiers.
Apart from the sharing, the MTL baseline is identical to the simple baseline.

\paragraph{Random Baseline.}
Finally, we provide a simple random baseline by sampling the outcome vectors from discrete uniform distribution. The random baseline is an average performance over 100 instantiations of this baseline.
\looseness=-1

\section{Experimental Setup}
\paragraph{Pre-trained Language Models.}\label{para:plm} 
We obtain high-dimensional representations $\enc(\facts)$ by fine-tuning one of the following pre-trained language models with $\facts$ as an input:
\begin{itemize}
    \item We first consider $\BERT$ because it is a widely used model in legal AI \cite{lexGLUE}.
    \item Second, we consider $\LBERT$, because it is trained on legal text, which should give it an advantage in our setting.\looseness=-1
    \item Finally we use the $\longformer$ model. $\longformer$ is built on the same $\transformer$ \cite{vaswani2017attention} architecture as $\BERT$ \cite{devlin-etal-2019-bert} and $\LBERT$ \cite{chalkidis-etal-2020-legal}, but it can process up to $4{,}096$ tokens. 
We select this architecture because the facts of legal documents often exceed $512$ tokens; a model that can process longer documents could therefore be better suited to our needs.
\end{itemize}

\looseness=-1

\paragraph{Training Details.} 
All our models are trained with a batch size of $16$. We conduct hyperparameter optimisation across learning rate $\{3\mathrm{e}{-4}, 3\mathrm{e}{-5}, 3\mathrm{e}{-6}\}$, dropout $\{0.2, 0.3, 0.4\}$ and hidden size of $\{50, 100, 200, 300\}$.
We truncate individual case facts to $512$ tokens for $\BERT$ and $\LBERT$ or $4{,}096$ tokens for the $\longformer$. 
Our models are implemented using the PyTorch \cite{paszke2019pytorch} and Hugginface \cite{Wolf2019HuggingFacesTS} libraries. 
We use Adam for optimization \cite{DBLP:journals/corr/KingmaB14} and train all our models on $1$ Tesla V$100$ $32$GiB GPU's for a maximum of $1$ hour. 
We train for a maximum of $10$ epochs.\footnote{All our code is available on \href{https://github.com/valvoda/Negative-Precedent-in-Legal-Outcome-Prediction}{Github}.}
The models are trained on the training set, see \cref{tab:splits}. 
We report the results on the test set for the models that have achieved the lowest loss on the validation set.
\looseness-1

\begin{table}[t]
\centering
\begin{tabular}{lcccc}
\toprule
\multicolumn{4}{c}{\textbf{\citeauthor{chalkidis-etal-2021-paragraph} Corpus}} \\
\midrule
\textbf{Outcome} & \textbf{Train} & \textbf{Validation} & \textbf{Test} \\
\midrule
Positive & 8046  & 835  & 851  \\
Negative & 2259  & 279 & 289  \\
Claims & 8836  & 985 & 991  \\
\midrule
\multicolumn{4}{c}{\textbf{Outcome Corpus}} \\
\midrule
\textbf{Outcome} & \textbf{Train} & \textbf{Validation} & \textbf{Test} \\
\midrule
Positive & 7542   & 844  & 925  \\
Negative & 4413   & 498 & 560  \\
Claims & 8372  & 931 & 1034  \\
\bottomrule
\end{tabular}
 \caption{Number of cases with at least one positive or negative outcome label in the dataset.}
  \label{tab:splits}
\end{table}

\section{Legal Corpora}\label{sec:corpora}

We work with the ECtHR corpus,\footnote{While ECtHR interacts with civil law jurisdictions, its judges rely on precedent \cite{valvoda-etal-2021-precedent}.} which contains thousands of instances of case law pertaining to the European Convention of Human rights (ECHR). ECtHR cases contain a written description of case facts, which is our $\facts$, and information about claims and outcomes. Since positive outcomes are a subset of all claims, the exclusion set of claims and positive outcomes constitutes the set of negative outcomes.

\paragraph{\chalkidis .} 
To obtain the golden labels for outcomes and claims we first rely on the \citet{chalkidis-etal-2021-paragraph} scrape of the ECHR corpus that contains \emph{alleged violations} and \emph{violations} labels. The violations are case outcomes, while the alleged violations are the main claims of the case.\looseness-1

\paragraph{\precedent .} 
Since violations are only the main claims of the case, to investigate the full set of claims (and negative outcomes) we process the \citet{chalkidis-etal-2019-neural} scrape of the online HUDOC\footnote{See \cref{appendix:glossary} for examples from our dataset or \href{https://hudoc.echr.coe.int/}{HUDOC} for all the ECHR caselaw.} database and extract the full set of claims using regular expressions.\looseness-1

We conduct all our experiments on both corpora.
However, not all of the Articles of ECHR are interesting from the perspective of a legal outcome, since not all of them can be claimed by a lawyer.
Out of the $51$ Articles of the convention, only Articles $2$ to $18$ contain the rights and freedoms, whereas the remaining Articles pertain to the court and its operation. 
The rights and freedoms are what a lawyer can claim, the focus of our work.
We therefore restrict our study to predicting the outcome of these core rights.
Furthermore, we remove any Articles that do not appear in the validation and test sets.
This leaves us with $K = 17$ and $K = 14$ for the \chalkidis{} and \precedent{} respectively.
\cref{tab:splits} shows the number of cases containing negative outcome vs positive outcome across the training/validation/test splits.
The full distribution of Articles over cases in both corpora can be found in \cref{appendix:corpora}.
\looseness -1

\begin{figure*}[h]
\subfloat[Negative outcome results on the \corpus. \label{fig:res_precedent_neg}]
  {\includegraphics[width=.5\linewidth]{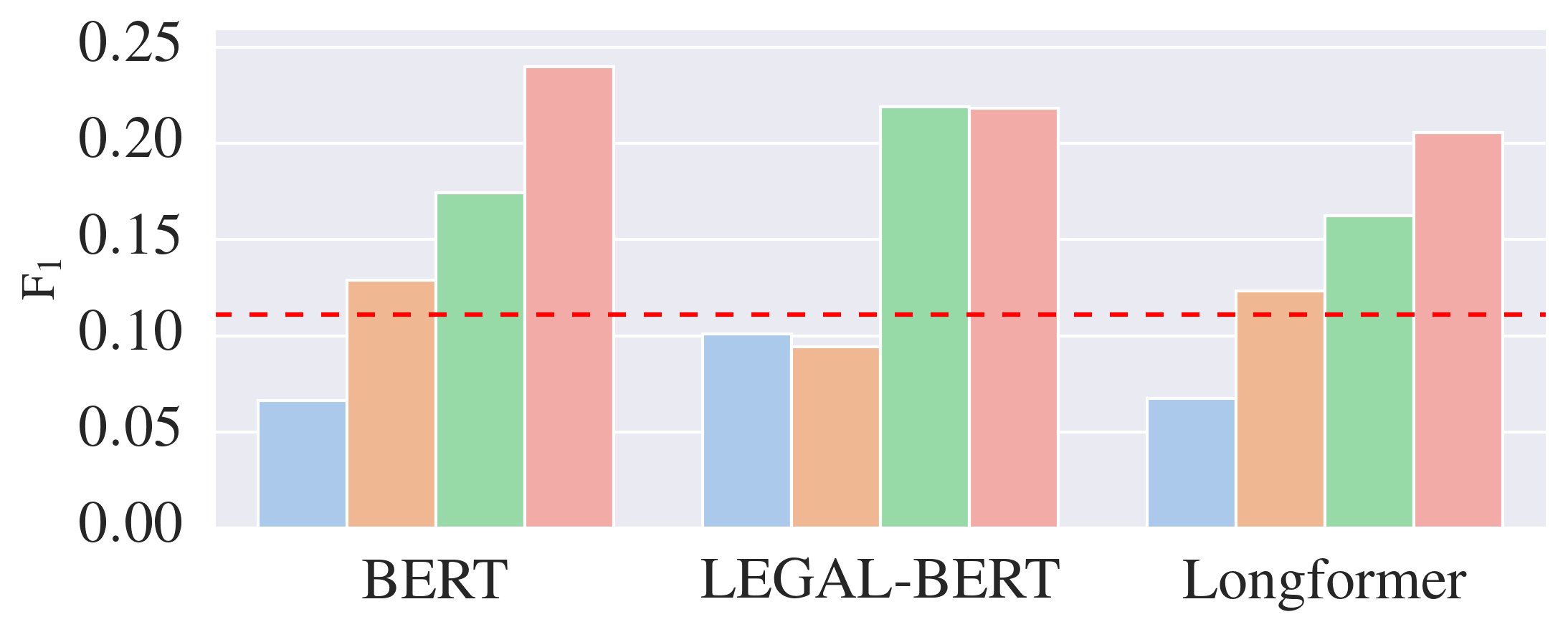}}\hfill
\subfloat[Negative outcome results on the \citeauthor{chalkidis-etal-2021-paragraph} corpus. \label{fig:res_alleged_neg}]
  {\includegraphics[width=.5\linewidth]{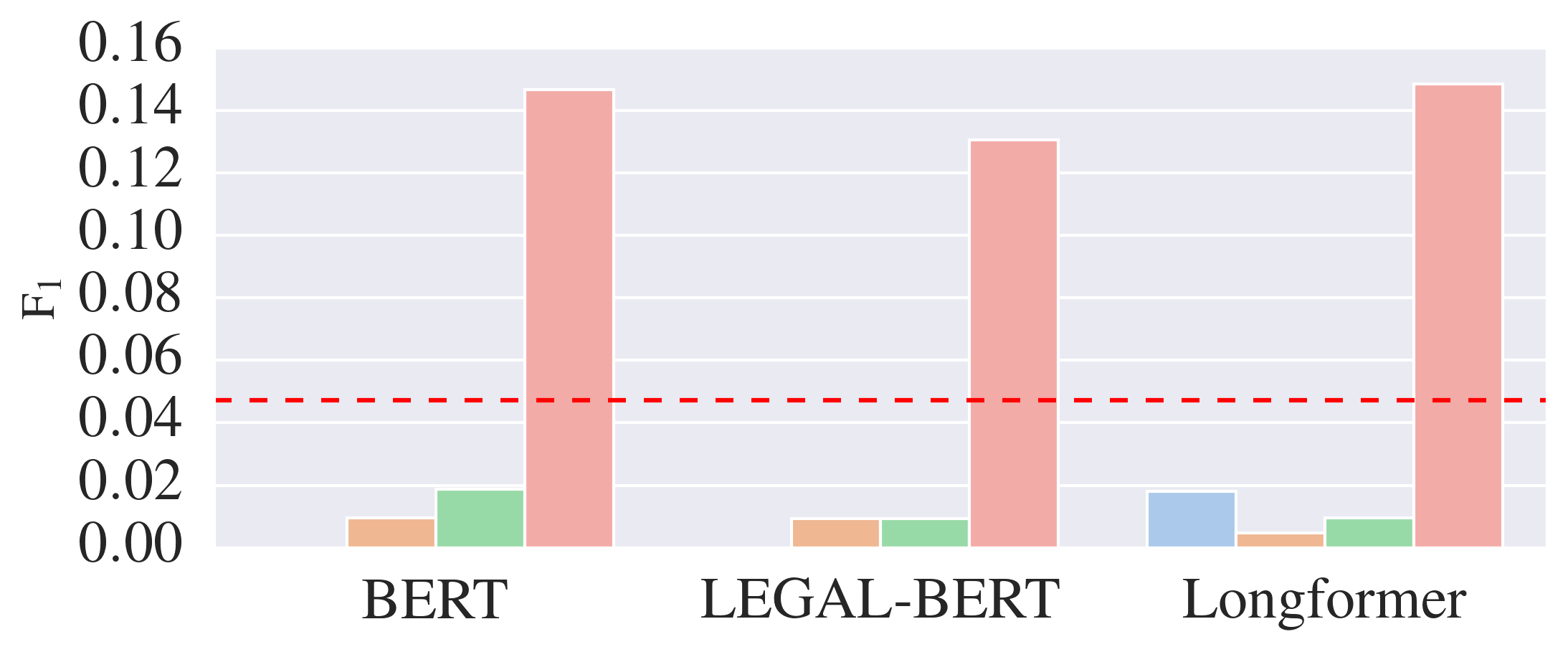}}\hfill
\caption{Results for  \textbf{\textcolor{myblue}{Simple}},  \textbf{\textcolor{myorange}{MTL}}, \textbf{\textcolor{mygreen}{Joint}} and  \textbf{\textcolor{myred}{Claim-Outcome}} models. Dashed \textcolor{red}{line} is our random baseline.}\label{fig:results}
\end{figure*}

\begin{table*}[]
\centering
\begin{tabular}{llrrrrrrrr}
                                                                                     & \textbf{}    & \multicolumn{4}{c}{\textbf{Outcome Corpus}}                                                                                              & \multicolumn{4}{c}{\textbf{\citeauthor{chalkidis-etal-2021-paragraph} Corpus}}                                                                                               \\
\toprule
\textbf{Model}                                                                       & \textbf{LLM} & \multicolumn{1}{l}{\textbf{Pos}} & \multicolumn{1}{l}{\textbf{Neg}} & \multicolumn{1}{l}{\textbf{Null}} & \multicolumn{1}{l}{\textbf{All}} & \multicolumn{1}{l}{\textbf{Pos}} & \multicolumn{1}{l}{\textbf{Neg}} & \multicolumn{1}{l}{\textbf{Null}} & \multicolumn{1}{l}{\textbf{All}} \\
\midrule
\multirow{3}{*}{\textbf{Claim-Outcome}}                                                  & $\BERT$         & 74.80                            & \textbf{24.01}                   & 95.53                             & 64.78                            & 63.85                            & 14.65                            & 97.15                             & \textbf{58.55}                   \\
                                                                                     & $\LBERT$         & 74.90                            & 21.83                            & 95.49                             & 64.07                            & 64.47                            & 13.05                            & 97.14                             & 58.22                            \\
                                                                                     & $\longformer$   & 74.23                            & 20.55                            & 95.17                             & 63.32                            & 63.53                            & \textbf{14.84}                   & 97.21                             & 58.53                            \\
\midrule
\multirow{3}{*}{\textbf{Joint}}                                                      & $\BERT$          & 76.24                            & 17.43                            & 95.46                             & 63.04                            & 65.15                            & 1.87                             & 97.07                             & 54.70                            \\
                                                                                     & $\LBERT$        & 76.96                            & 21.93                            & \textbf{95.71}                    & \textbf{64.87}                   & \textbf{67.08}                   & 0.94                             & \textbf{97.19}                    & 55.07                            \\
                                                                                     & $\longformer$    & \textbf{77.15}                   & 16.24                            & 95.49                             & 62.96                            & 65.94                            & 0.95                             & 97.11                             & 54.67                            \\
\midrule
\multirow{3}{*}{\textbf{\begin{tabular}[c]{@{}l@{}}MTL \\ Baseline\end{tabular}}}    & $\BERT$         & 75.75                            & 12.90                            & -                               & -                              & 63.21                            & 0.95                             & -                               & -                              \\
                                                                                     & $\LBERT$       & 76.73                            & 9.44                             & -                               & -                              & 65.00                            & 0.95                             & -                               & -                              \\
                                                                                     & $\longformer$   & 75.83                            & 12.34                            & -                               & -                              & 63.36                            & 0.47                             & -                               & -                              \\
\midrule
\multirow{3}{*}{\textbf{\begin{tabular}[c]{@{}l@{}}Simple \\ Baseline\end{tabular}}} & $\BERT$        & 75.06                            & 6.62                             & -                               & -                              & 65.04                            & 0.00                             & -                               & -                              \\
                                                                                     & $\LBERT$       & 74.85                            & 10.09                            & -                               & -                              & 65.51                            & 0.00                             & -                               & -                              \\
                                                                                     & $\longformer$    & 74.12                            & 6.72                             & -                               & -                              & 63.92                            & 1.81                             & -                               & -                         \\
\bottomrule
\end{tabular}
\caption{$F_1$-micro averaged scores for all the models considered over the two datasets.}
\label{table:results}
\end{table*}

\section{Results}
Following \citet{chalkidis-etal-2019-neural}, we report all results as $F_1$ micro-averaged. We report significance using the two tailed paired permutation tests with $p < 0.05$. The bulk of our results is contained in \cref{table:results}. We report individual conclusions in the following paragraphs:

\paragraph{Negative outcome prediction is challenging.} First, we compare the positive and negative outcome prediction performance on our outcome corpus and find that while the best simple baseline model achieves $75.06$ $F_1$ on positive outcomes, the same model achieves only $10.09$ $F_1$ on negative outcomes.
In fact, the model fails to beat our random baseline of $11.12$ $F_1$ on negative outcomes.
The same trend holds over all our model architectures, all the underlying language models and both datasets under consideration. Every time, the negative outcome performance is significantly lower than that of positive outcomes.
Therefore, our first conclusion is that negative outcome is simply harder to predict than its positive counterpart.
\looseness-1

\paragraph{Claim-outcome model improves negative outcome prediction.}
We observe a large and significant improvements using our \model{} on the task of negative outcome prediction; see \cref{fig:res_precedent_neg} and \cref{fig:res_alleged_neg}. 
Our \model{} is better than every baseline model under consideration, a finding that holds over three underlying language models and both datasets.
A single exception to this rule is the joint model, which insignificantly beats our \model{} (by $0.1$) on the \corpus{} using the $\LBERT$ LLM.
Overall, where the best \model{} achieves $24.01$ $F_1$ on the \corpus{} and  $14.84$ $F_1$ on the \citeauthor{chalkidis-etal-2021-paragraph} corpus, the best simple baseline model only achieves $10.09$ and $1.81$ $F_1$, respectively.
Therefore, our second conclusion is that enforcing the relationship between claims and outcomes improves negative outcome prediction.
We expand our discussion on this in \cref{app:baselines}.\looseness-1

\paragraph{Joint-model improves positive outcome prediction.}
Turning to the positive outcome prediction task, we see that the simple baseline and our \model{} have a comparable $F_1$.
The joint model on the other hand improves over either baseline and achieves the best $F_1$ on both the \corpus{} and the \citeauthor{chalkidis-etal-2021-paragraph} corpus ($77.15$ and $67.08$ $F_1$, respectively).
Since the simple baseline model using pre-trained $\BERT$ is the state-of-the-art model for positive outcome prediction \cite{lexGLUE}, our third conclusion is that jointly training on positive and negative outcomes is a better way of learning how to predict a positive outcome of a case.
\looseness-1

\paragraph{Impact of pre-training.}
In line with the standard results on LexGLUE task A \cite{lexGLUE},
we find that $\LBERT$ and $\longformer$ fail to consistently outperform $\BERT$. 
None the less, $\LBERT$ has a significant positive effect on negative outcome prediction for the joint model. 
It improves over $\BERT$ ($17.43$ $F_1$) and $\longformer$ ($16.24$ $F_1$) based models and achieves $21.93$ $F_1$.
It is also the underlying language model for the highest performing model for the outcome corpus (achieving $64.87$ $F_1$ overall).
Meanwhile, $\longformer$ sets the highest positive outcome performance on the same corpus ($77.15$ $F_1$).
We therefore find both longer document encoding and legal language pre-training useful in certain narrow settings, although it seems that the choice of model architecture has a larger effect on the performance than the choice of the language model size or pre-training material.

\paragraph{Which model is the best?} Finally, we turn to the question of what is the best model of outcome prediction; the joint model or the \model{}?
Towards answering this question we take an average $F_1$ over all three random variables under consideration; the best model of outcome should do well at distinguishing between positive, negative and null outcome.
We find that while the joint model has an insignificant edge over the \model{} on the \corpus{} (by $0.1$), on the \citeauthor{chalkidis-etal-2021-paragraph} corpus the \model{} significantly improves over the joint model (by $3.48$ $F_1$).
This leads us to believe that \model{} is overall the better model for legal outcome prediction.
However, both models are valuable in their own right.
Where the joint model improves over state-of-the-art positive outcome prediction models, the \model{} doubles their performance on the negative outcome task.

\section{Discussion}\label{sec:discussion}

\begin{figure}[t]
\centering
\includegraphics[width=7.9cm]{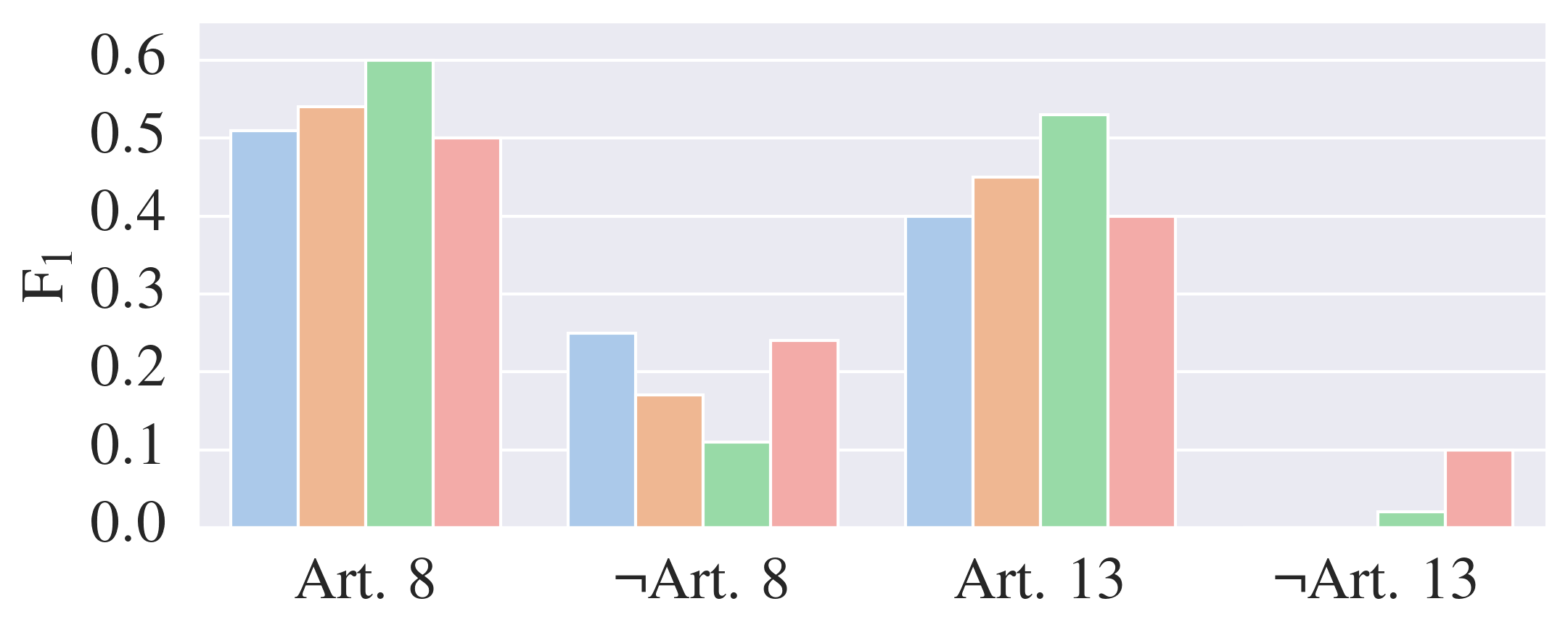}
\caption{Article 8 and 13 results for \textbf{\textcolor{myblue}{Simple}},  \textbf{\textcolor{myorange}{MTL}}, \textbf{\textcolor{mygreen}{Joint}} and  \textbf{\textcolor{myred}{Precedent}} models.}
\label{fig:subset}
\end{figure}

The results reported above raise the question of why models severely underperform on negative outcome prediction.
The simplest answer could be the amount of training data that is available for each task.
We test this hypothesis by comparing the performance on Articles $8$ ($796$ negative vs. $654$ positive examples) and $13$ ($1197$ negative vs. $1031$ positive examples) of ECHR in our \corpus{}, where there is more training data available for the negative outcome than for the positive outcome.
The results, given in \cref{fig:subset}, show that even when the model has more training data for negative outcome than the positive outcome, predicting negative outcome is still harder.
In particular for Article $13$, the amount of training data is higher than for Article $8$, yet the drop in performance between positive and negative outcome prediction is still dramatic.
In fact, while the scores achieved by the \model{} are still low, the other models (except our joint model) fail to predict a single negative outcome correctly for Article $13$.
We therefore believe that the performance drop is likely to be more related with the complexity of the underlying task, than with the imbalance of the underlying datasets.
\looseness-1

To find a better explanation of the performance asymmetry, we now turn in our discussion to the legal perspective.
In precedential jurisdictions, of which ECtHR is one \cite{zupancic, lupu2010role, valvoda-etal-2021-precedent}, the decisions of a case are binding on future decisions of the court. 
Two cases with the same facts should therefore arrive at the same outcome. 
Of course, in reality, the facts are never the same. 
Rather, cases with similar circumstances will, broadly speaking, lead to similar outcomes. 
This is achieved by \emph{applying} the precedent. 
In such cases, the judge will in effect say that the new case is not substantially different from an already existing case and therefore the same outcome will be propagated. \looseness-1

On the other hand, if the previous precedent is not to be followed, the judge needs to \emph{distinguish} the case at hand from the precedent.
Distinguishing the case from the precedent is a more involved task than applying the precedent.
It requires identifying what exactly about the new facts sets the new case apart from the previous one. 
This can of course be done for both cases with positive and negative outcome.
Both can be applied or distinguished.\footnote{Overruling is another option, though it is exceptionally rare at the ECtHR \cite{dzehtsiarou2017law}.} 
Since judges deal with claims, each of which comes with an argument built around the precedent that favours the claimant's viewpoint, we believe that negative outcomes overwhelmingly rely on distinguishing the case from the precedent. 
This is evidenced in the yearly reports of the \citet{ecthr_report}, which list cases where the judges decided to distinguish the facts of the case at hand.
Distinguishing almost always leads to a negative outcome. 
We observe the same trend in our ECtHR corpus.
\looseness-1 

It might therefore be the case that while there is such a thing as a prototypical positive precedent, there is no prototypical negative precedent.
This could explain why the simpler architectures struggle to learn to predict it.
While a simpler model is ill-suited for the task since it is trained to find a similarity between the negative outcome cases, our claim-outcome model does not assume that negative outcome cases are similar in the first place.
Instead, our model assumes similarity between claims.
Since claim prediction can be modelled with a high accuracy \cite{chalkidis-etal-2021-paragraph}, we can reveal the negative outcome as a disagreement between a judge and a lawyer (i.e. claims and the outcomes).\looseness-1

By investigating individual cases in \chalkidis, we can  identify a further possible explanation for the baseline model performance.
For instance, 
the case of \href{https://hudoc.echr.coe.int/fre?i=001-181583}{Wetjen and Others v. Germany} (Wetjen)
is concerned with Article 8: Right to respect for private and family life.
In this case, religious parents used caning (among other methods) as a punishment for their children. 
The German State intervened and placed the children in foster care. 
The parents claim interference with their right to family life.
On a superficial level, the case is similar to two Article 8 cases  both cited in Wetjen: that of \href{https://hudoc.echr.coe.int/eng?i=001-99817}{Shuruk v. Switzerland} (Shuruk), and \href{https://hudoc.echr.coe.int/eng?i=001-70957}{Suss v. Germany} (Suss).

In Shuruk, religious parents fight for an extradition of a child. The mother of the child argues that it would be an interference with her right to a family life if the child was to be extradited to the husband, who has joined an ultra-orthodox Jewish movement.
A component of the case is an allegation of domestic violence the husband was supposed to have perpetrated against his wife.
In Suss, the German State has denied a divorced father access to his daughter due to the frequent quarrels between the parents during the visits. 
The father alleges breach of Article 8.
In Wetjen and Suss, the judges have decided a violation of Article 8 has not occurred, they have ruled the opposite in Shuruk.\looseness-1

On the surface, the facts are alike, especially between Shuruk and Wetjen -- both cases contain elements of abuse, religion and state intervention.
However, to a human lawyer, the distinction between the cases is fairly trivial.
In Wetjen, the State is allowed to intervene to protect a child from an abusive ultra-religious parents, which is very similar situation to Suss, where a State is allowed to intervene to protect the child from quarrels between divorced parents.\looseness-1

All our models are exposed to both Shuruk and Suss in the training set.
However, for the positive outcome baseline, the information about Suss being related to Article 8 is lost.
Conversely, for the negative outcome baseline, the information about Shuruk being related to Article 8 is lost.
It is therefore not surprising that the best performing negative\footnote{Simple Baseline $\longformer$.} and positive\footnote{Simple Baseline $\LBERT$.} outcome baseline models both get the Wetjen outcome prediction wrong.
On the other hand, the best claim--outcome model, which is trained to learn that both Shuruk and Suss are related to Article 8 via the claim prediction objective, makes the correct outcome prediction in the Wetjen case.\looseness-1

In conclusion, our claim-outcome model is indeed a better way of modelling negative outcome, but its superiority  is not due to the fact that it is  learning anything about the law itself.
It simply leverages the fact that positive outcomes and claims are easy to predict and enforces the relationship between them.
To identify the negative outcome with high $F_1$ will require deeper understanding of law than our models are currently capable of.
\looseness-1

\section{Related Work}
Juris-informatics can trace its origins all the way to the late 1950s \citep{kort_1957, nagel1963applying}.
The pioneers used rule-based systems to successfully capture aspects of legal reasoning, using thousands of hand-crafted rules \citep{hypo}. 
Yet due to the ever-changing rules of law, these systems were too brittle to be employed in practice. 
Particularly in common law countries, the majority of law is contained in case law, where cases are transcripts of the judicial decisions. 
This allows the law to constantly change in reaction to each new decision. 
The advances of natural language processing (NLP) in the past two decades have rejuvinated the interest in developing applications for the legal domain.
Areas explored include question answering \citep{monroy09}, legal entity recognition \citep{cardellino-etal-2017-legal}, text summarisation \citep{Hachey}, judgement prediction \citep{xu2020distinguish}, majority opinion prediction \cite{Valvoda2018} and \emph{ratio decidendi} extraction \cite{valvoda}.

Our work is similar to the recent study of Chinese law judgement prediction by \citet{zhong-etal-2018-legal} and  \citet{xu2020distinguish}, who break down court judgements into the applicable law, charges and terms of penalty. 
Operating in the civil law system (which outside of China is also used in Germany, and France, inter alia), they argue that predicting applicable law is a fundamental subtask, which will guide the prediction for other subtasks. 
In the context of ECHR law, we argue that legal claims are one such guiding element for outcome prediction. 
While similar, applicable law and claims are different. 
In the work above, the judge selects the applicable law from the facts as part of reaching the outcome. 
This is not the case for ECHR law, or any other precedential legal system known to the authors, where the breach of law is claimed by a lawyer, not a judge.

Finally, the ECtHR dataset has been collected by \citet{chalkidis-etal-2019-neural}, who have predicted outcomes of the ECHR law and the corresponding Articles using neural architectures. Our work builds on their research by reinstantiating the outcome prediction task on this dataset to include negative precedent. Similar datasets, which one could apply our method to, include Caselaw Access Project and US Supreme Court caselaw.\footnote{See \href{https://www.oyez.org/cases/2018}{US Supreme Court corpus} and \href{https://case.law/}{Caselaw Access Project}.} 

\looseness-1

\section{Conclusion}
While positive and negative outcomes are equally important from the legal perspective, the current legal AI research has neglected the latter.
Our findings suggest that negative outcome is much harder to predict than positive outcome, at least for current deep learning models.
This has severe implications for how well the current legal models can model judicial outcome.
The same models that predict positive outcome with $75.06$ $F_1$ fall short of a random baseline of $10.09$ $F_1$ on the negative outcome prediction task.
\looseness -1

We discuss possible reasons why negative outcome prediction is so much harder to learn.
Specifically, we suspect that negative outcomes are mostly caused by a judge distinguishing the case from its precedent. 
This lead us to believe that learning to predict negative outcomes requires more legal understanding than the current models are capable of.  We believe that negative outcome prediction is therefore a particularly attractive task for evaluating progress in legal AI.

Our work improves over the existing models by inducing the relationship between the judge and the lawyer in our \model{} architecture. However, the best negative outcome prediction model achieves only a third of the performance of the positive outcome one.
In future work we hope to study the phenomenon of precedent more closely, with the aim of building models capable of narrowing this performance gap.
One possible avenue would be to relax our \cref{assumption1} to study the potential relationships between the individual legal Articles. We leave this direction for future work.

\section{Ethical Considerations}\label{sec:ethics}

Legal models similar to the ones we study above have been deployed in the real world.
Known applications include risk assessment of detainees in the US \cite{Kehl2017AlgorithmsIT} and sentencing of criminals in China \cite{Xiao2018CAIL2018AL}. 
We believe these are not ethical use cases for legal AI. 
One must not be tempted to think of outcome prediction as equivalent to some medical task, such as cancer detection, with a breach of law seen as a `tumour' that is either there or not. 
This is a naive viewpoint that ignores the fact that the legal system is a human construct in which the decision makers, the judges, play a role that can shift the truth, something that is impossible when it comes to natural laws. \looseness-1

Herein lies the ethical dubiousness of any attempt at modelling judges using AI. 
Unlike in the domain of medicine, where identifying the underlying truth is essential for treatment, and thus a successful machine diagnostician is in theory a competition for the human one, in the domain of law the validity of the decision is poised solely on the best intentions of the judge. 
For some judges this pursuit of the `right' outcome can go as far as defiance of legal precedent. 
We therefore argue a judge should not be replaced by a machine and caution against the use of our, or any other legal AI model currently available, towards automating the judicial system. \looseness-1

\appendix   

\section{Note on the Baselines}\label{app:baselines}

Comparing the MTL baseline and the joint model, one might come to the conclusion that there is no substantial difference between the models when it comes to predicting positive outcomes. 
While the joint model outperforms the MTL baseline on eleven out of the twelve experiments we test our models on, the improvement in performance on the positive outcome prediction over the outcome corpus is very narrow. 
However, there is an important difference between them.
The MTL baseline, much like the simple baseline, can predict positive and negative outcome simultaneously for the same Article.
This means that in our evaluation, the baseline models can cheat by predicting an Article to be simultaneously violated and not-violated.
This is another reason that the outcome prediction task needs to consider the legal relationship between positive and negative outcomes.
Ignoring the relationship of claims and outcomes makes both of our baselines fundamentally ill-suited for the task of outcome prediction. Hence, they are only useful for a comparison in our study.\looseness-1
\onecolumn

\section{Glossary \& Dataset Examples:}\label{appendix:glossary}

\begin{table}[h!]
  \centering
  \begin{tabular}{p{0.20\linewidth} p{0.7\linewidth}}
    \textbf{Legal Terms} \\
    \toprule
    \emph{Claim} & The allegation of a breach of law usually put forth by their legal counsel on behalf of the claimant.\\
    \midrule
    \emph{Positive Outcome} & Claims are assessed by judges in courts of law. If they think a claim is valid, they rule it as successful. The outcome of the case is a victory for the claimant; which we call the positive outcome in this paper.\\
    \midrule
    \emph{Negative Outcome} & On the other hand, the claimant can be unsuccessful in the court. The judge has decided against them in the court, in favour of the defendant, and we call this the negative outcome in this paper.\\
    \midrule
    \emph{Facts} & The description of what happened to the claimant. This includes more general descriptions of who they are, circumstances of the perceived violation of their rights and the proceedings in domestic courts before their appeal to ECtHR.\\
    \midrule
    \emph{Precedent} & Cases that have been cited by the judges as part of their arguments. \\
    \midrule
    \emph{Binding} & Judges are expected to adhere to the binding rules of law and decide future access accordingly. \\
    \midrule
    \emph{\begin{tabular}{@{}l@{}} Stare Decisis\end{tabular}} & New cases with the same facts to the already decided case should lead to the same outcome. This is the doctrine of precedent by which judges can create law. \\
    \midrule
    \emph{Caselaw} & Transcripts of the court proceedings. \\
    \midrule
    \emph{ECHR} & European Convention of Human Rights, comprises of the Convention and the Protocols to the convention. The Protocols are the additions and amendments to the Convention introduced after the signing of the original Convention.\\
    \midrule
    \emph{ECtHR} & European Court of Human Rights, adjudicates ECHR cases. \\
    \midrule
    \emph{Apply} & A judge applies the precedent when she decides on the outcome of a case via an analogy to an already existing case. \\
    \midrule
    \emph{Distinguish} & Conversely, a judge distinguishes the case from the already existing cases when she believes they are not analogous. \\
    \bottomrule
  \end{tabular}
  \label{glossary:1}
\end{table}

\begin{table}[h!]
  \centering
  \begin{tabular}{p{0.555\linewidth} p{0.07\linewidth} p{0.11\linewidth} p{0.11\linewidth}}
    \textbf{ECtHR Example} & & & \\
    \toprule
    \textbf{Facts} & \textbf{Claims} & \textbf{Positive Outcomes} & \textbf{Negative Outcomes} \\
    \midrule
    ``Ms Ivana Dvořáčková was born in 1981 with Down Syndrome (trisomy 21) and a damaged heart and lungs. She was in the care of a specialised health institution in Bratislava. In 1986 she was examined in the Centre of Paediatric Cardiology in Prague‑Motole where it was established that, due to post-natal pathological developments, her heart chamber defect could no longer be remedied...'' for more see \href{http://hudoc.echr.coe.int/eng?i=001-93768}{Case of Dvoracek and Dvorackova v. Slovakia} & Articles: 2, 6, 8, 14 & Articles: 2, 6 &  Articles: 8, 14 \\
    \bottomrule
  \end{tabular}
  \label{glossary:2}
\end{table}

\section{Corpora Statistics:}\label{appendix:corpora}

\begin{figure}[hbt]
\centering
\includegraphics[width=15.7cm]{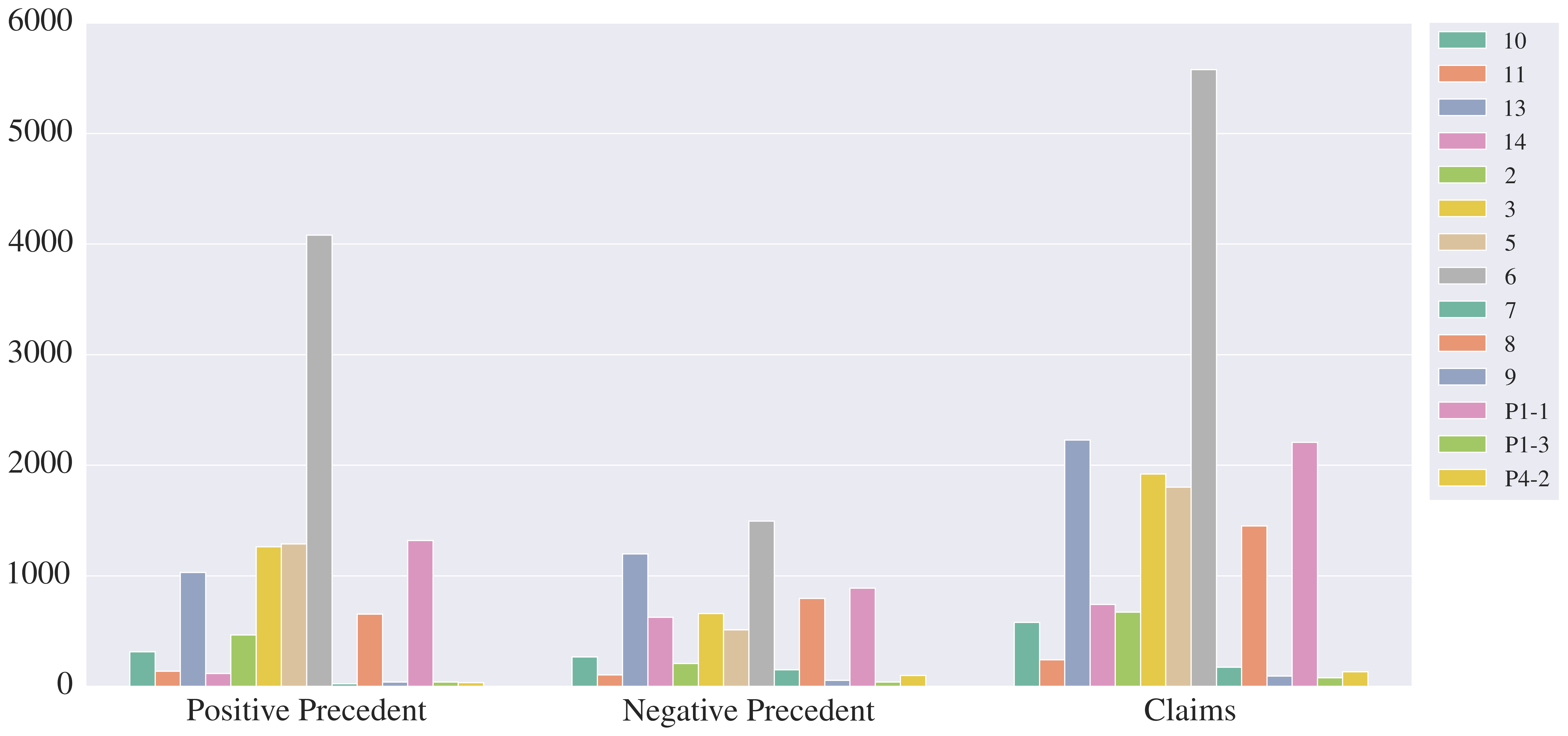}
\caption{Distribution of Articles over training data in our outcome corpus.}
\label{fig:precedent_stats}
\end{figure}

\begin{figure}[hbt]
\centering
\includegraphics[width=15.7cm]{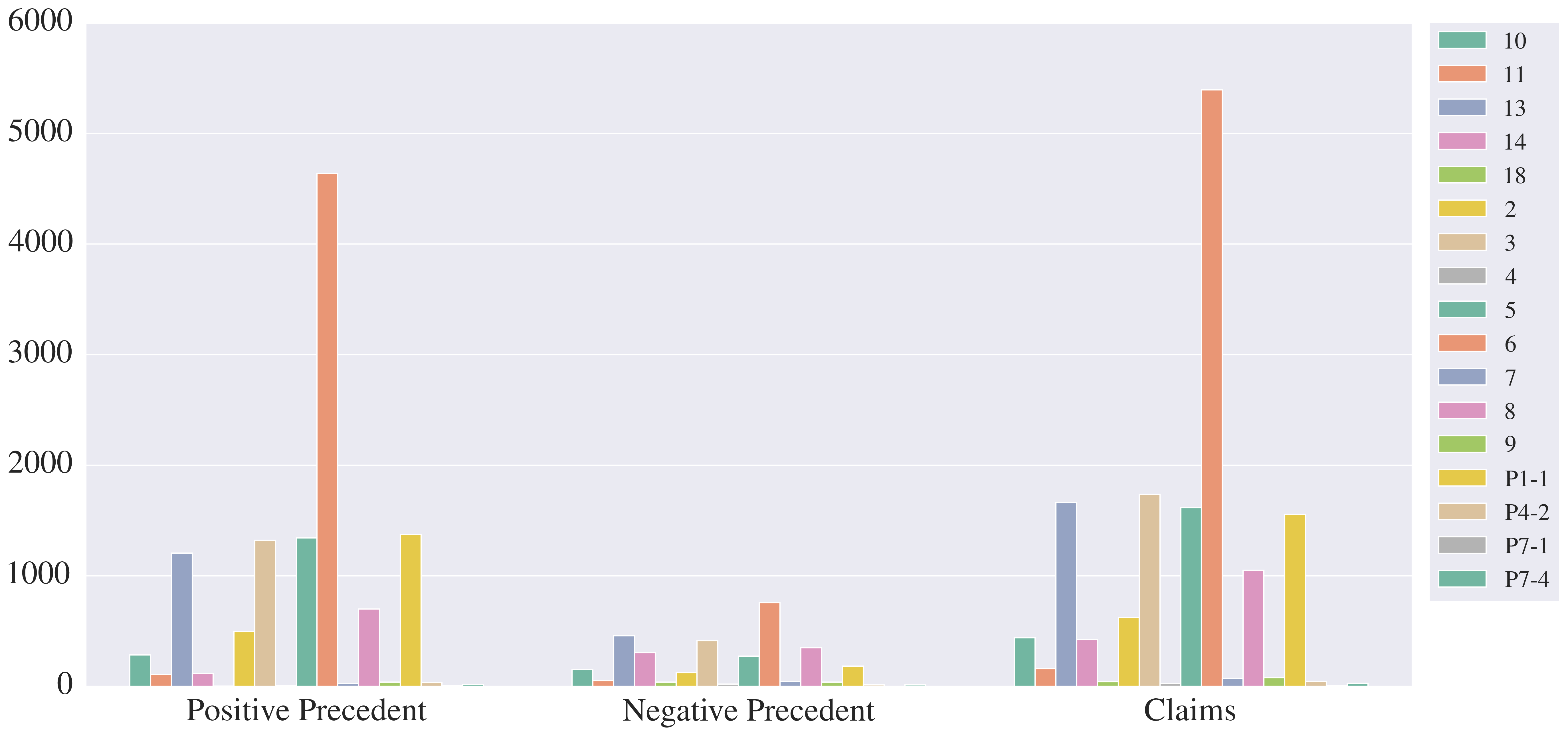}
\caption{Distribution of Articles over training data in \citeauthor{chalkidis-etal-2021-paragraph} corpus.}
\label{fig:alleged_stats}
\end{figure}

\twocolumn

\bibliography{anthology.bib,custom.bib}
\bibliographystyle{acl_natbib}




\end{document}

